\documentclass[10pt,letter]{IEEEtran}
\usepackage{graphicx}
\usepackage{float}
\usepackage{eurosym}

\usepackage[T1]{fontenc}
\begin{document}


\title{
    A Low Cost Discrete Digital Isolator Circuit.
}
\author{
Thomas Conway
\thanks{
Thomas Conway is an Associate Professor in the ECE dept. of the University of Limerick, Limerick, Ireland
}

\thanks{Contact:
Dr. Thomas Conway
Lecturer, ECE Dept
University of Limerick
National Technology Park
Limerick, Ireland.
Tel +353 61 202628,
Email thomas.conway@ul.ie
}
}

\maketitle

\begin{abstract}

    This work presents a fully discrete, low‑cost digital isolator requiring no specialized ICs and implemented entirely with general‑purpose transistors and a two‑layer PCB‑embedded air‑core transformer. The design avoids vendor lock‑in and long‑term component obsolescence risks, while providing >1 kV isolation, ~200 ns propagation delay, and validated NRZ data rates of 1 Mbps. A modified dual‑oscillator architecture enables inherent hardware lockout suitable for half‑bridge gate‑driver applications. Measured performance and PCB layout guidelines are provided.


\end{abstract}

\begin{IEEEkeywords}
      Galvanic Isolation, PCB Air Core Transformer, Discrete Circuit Implementation.
\end{IEEEkeywords}

\vspace{-0.2in}

\section{  Galvanic Digital Isolation}

  Galvanic isolation for digital signals is a widespread requirement
  and has been well served for a long time with opto-isolators and
  more recently integrated micro transformer and capacitive coupled
  commercial IC's \cite{ref:optoIso},\cite{ref:AnalogDevicesADuM361N},\cite{ref:TIcapiso}.

  However, for some applications such as low volume production or
  long lifetime products, a discrete solution using generic components
  may be desirable.  In these cases, a vendor specific IC or specialist
  process might lead to long lead times or issues with obsolescence.
  This issue of obsolescence or part discontinuation has been of concern
  for a long time in the electronics area \cite{ref:Obsolescence00},\cite{ref:Obsolescence25}.

  There is a need therefore in some applications for a discrete 
  digital isolation implementation that uses widely available components
  that have multiple sources and are likely to continue in production.
  Discrete general purpose transistors are one such example and the devices
  employed in this design have been in production for more that 40 years!

\section{Discrete Circuit Implementation}

  The circuit described here provides the digital isolation function 
  and consists of a simple, fast startup, discrete LC 
  high frequency oscillator using an air core inductor, that is 
  magnetically coupled to an identical air core inductor on the 
  isolated side and the induced voltage, is rectified to provide
  a binary digital output.

  The novelty of the circuit is the simplicity of the design
  and the use of ubiquitous general purpose transistors with the
  transformer (i.e. coupled inductors) simply implemented as
  PCB trace spiral coils on both sides of a standard PCB\cite{ref:pcbspiralInd}.

  The circuit simplicity allows the secondary and primary circuits
  to be implemented using surface mounted components on each
  respective PCB side without the need for any vias.  

  With 1.6mm thick FR4 PCB as used in this design, and the dielectric
  strength in excess of 20 kV/mm,  the breakdown voltages well into the
  several kV range is easily achieved.

  Thus high voltage isolation is delivered through the PCB material
  itself without the need for any high voltage components or concerns 
  over creep distance specifications\cite{ref:aircore2}\cite{ref:tiCreep}.

\begin{figure*}[htb]
\begin{center}
  \includegraphics[width=0.80\textwidth]{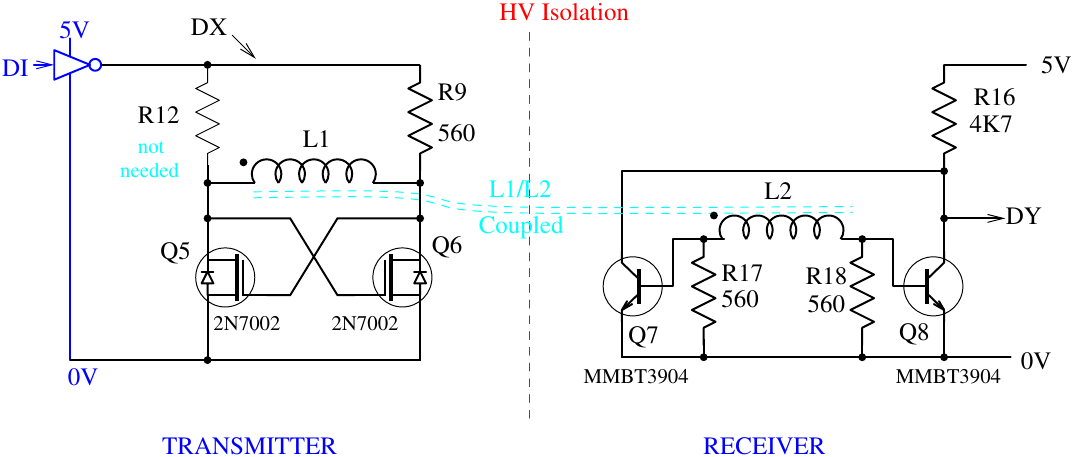}
\end  {center}
\caption{Schematic of Digital Isolator using Discrete Components}
\label{fig:basicschematic}
\end{figure*}

\subsection{Transmitter Side}
  Fig.~\ref{fig:basicschematic} shows the circuit schematic developed.  The oscillator is
  the well known cross coupled MOSFET pair which delivers a negative
  resistance between the two drain terminals.  An LC resonant circuit
  is normally connected across these two drain terminals, but in this
  case only an L (inductor) is used, with the parasitic capacitance
  of the drain source and gate source of the 2N7002 transistors (Q5,Q6)
  providing the resonance capacitance C.  Two drain resistors (R12,R9)
  provide the required operating current.  
  
  The circuit is driven by a
  5V logic signal and simply powers the oscillator to transmit a logic 1
  or leaves it un-powered, i.e. off for a logic 0.  When powered-on, a 
  high frequency oscillation excites the receiver circuit.  This is
  simply on-off keying in modulation terminology.
  
  Prototyping the transmitter circuit indicates a maximum reliable oscillation 
  frequency around 30 MHz with a minimum inductance of 0.4 uH.  Choosing
  a frequency of 15 MHz allows for margin in the design and reliable
  oscillation based on bench measurements with an inductance of 1.6uH.

  This level of inductance is readily achievable with air core PCB coils
  of manageable size\cite{ref:pcbspiralInd}.

  Breaking symmetry between the resistors R12, R9 results in a faster startup
  of the oscillator and in fact one of the resistors can be completely 
  omitted.

  \subsection{Receiver Side}
  The isolated receiver side is based on the identical inductor L2
  which is coupled to L1.  General purpose NPN transistors (MMBT3904s)
  Q7 and Q8 in conjunction with R17 and R18 effectively implement full
  wave rectification and amplification of the induced voltage across L2.  To see this,
  just consider Q7, L2 and R18 and R17, ignore Q8 which will be turned off.  
  With a positive
    voltage induced (of more than $2 \times V_{BE}$) in L2 (taking its 'dotted' side more positive)  
    and divided by the potential divider formed by R18 and R17,
  Q7 turns on, pulling its collector low and outputting a logic 0 at the
  output (Node DY).  R18 limits the base current into Q7 and the loading added
    onto the transmitter oscillator.  On the opposite 
    polarity of the high frequency cycle, Q8 (with R17, R18) performs identically 
  pulling the collector of Q4 low.
  During the zero crossings, the output does not go high due to the fact
  that the transistors Q7 and Q8 are bipolars and subject to saturation
  when turned on.  The slow recovery from saturation (typ. 100ns 
  for MMBT3904 at 1mA) effectively implements a low pass filter 
  function for the rectified signal and results in a stable logic signal 
  at node DY.

\section{PCB Implementation with Copper Trace Transformer}

  Using PCB traces for the coil windings and placing the transmitter
  part on opposite sides of the PCB provides for a convenient implementation
  \cite{ref:pcbspiralInd}.

\begin{figure}[htb]
\begin{center}
  \includegraphics[width=0.5\textwidth]{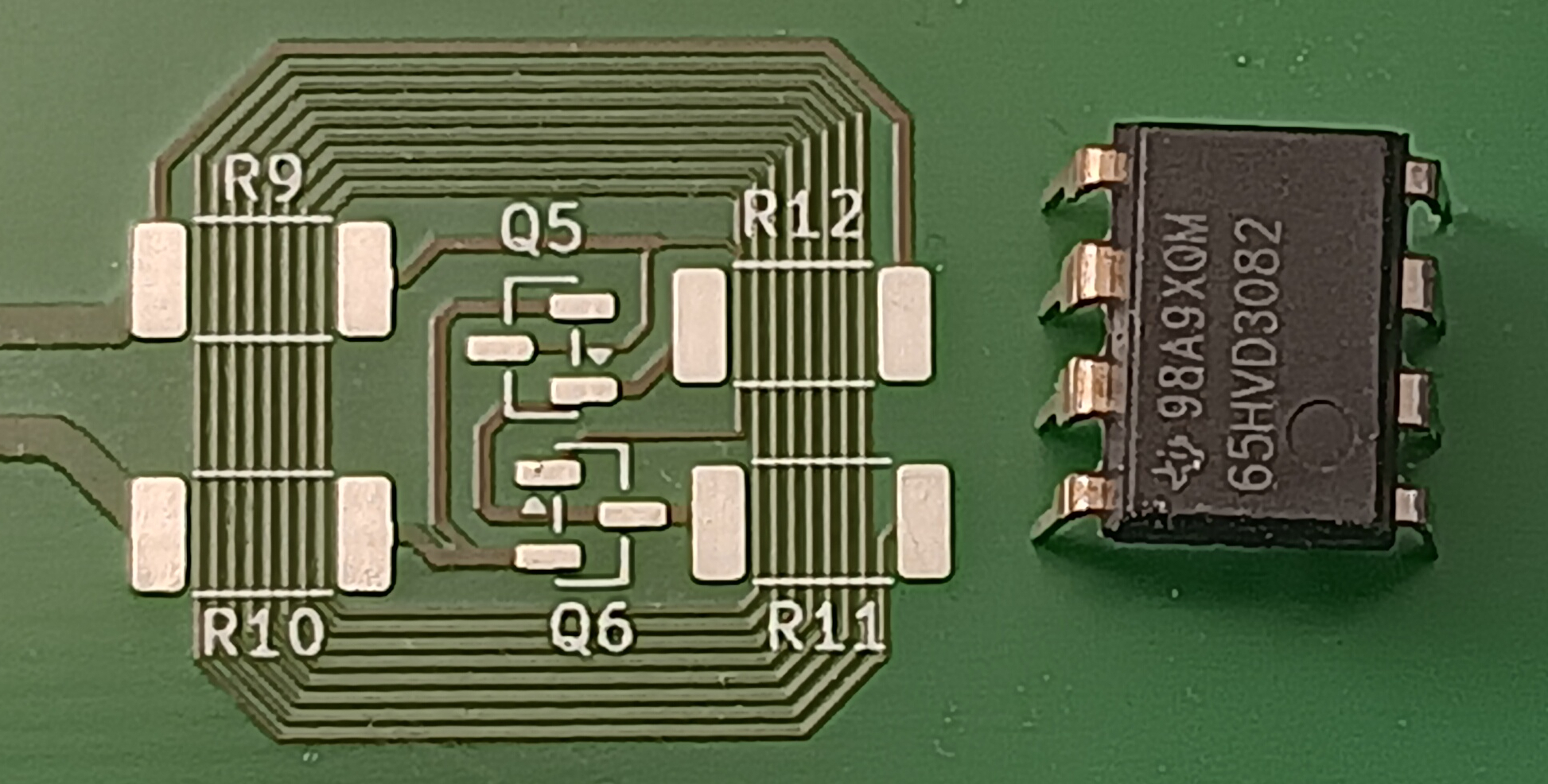}
  \includegraphics[width=0.5\textwidth]{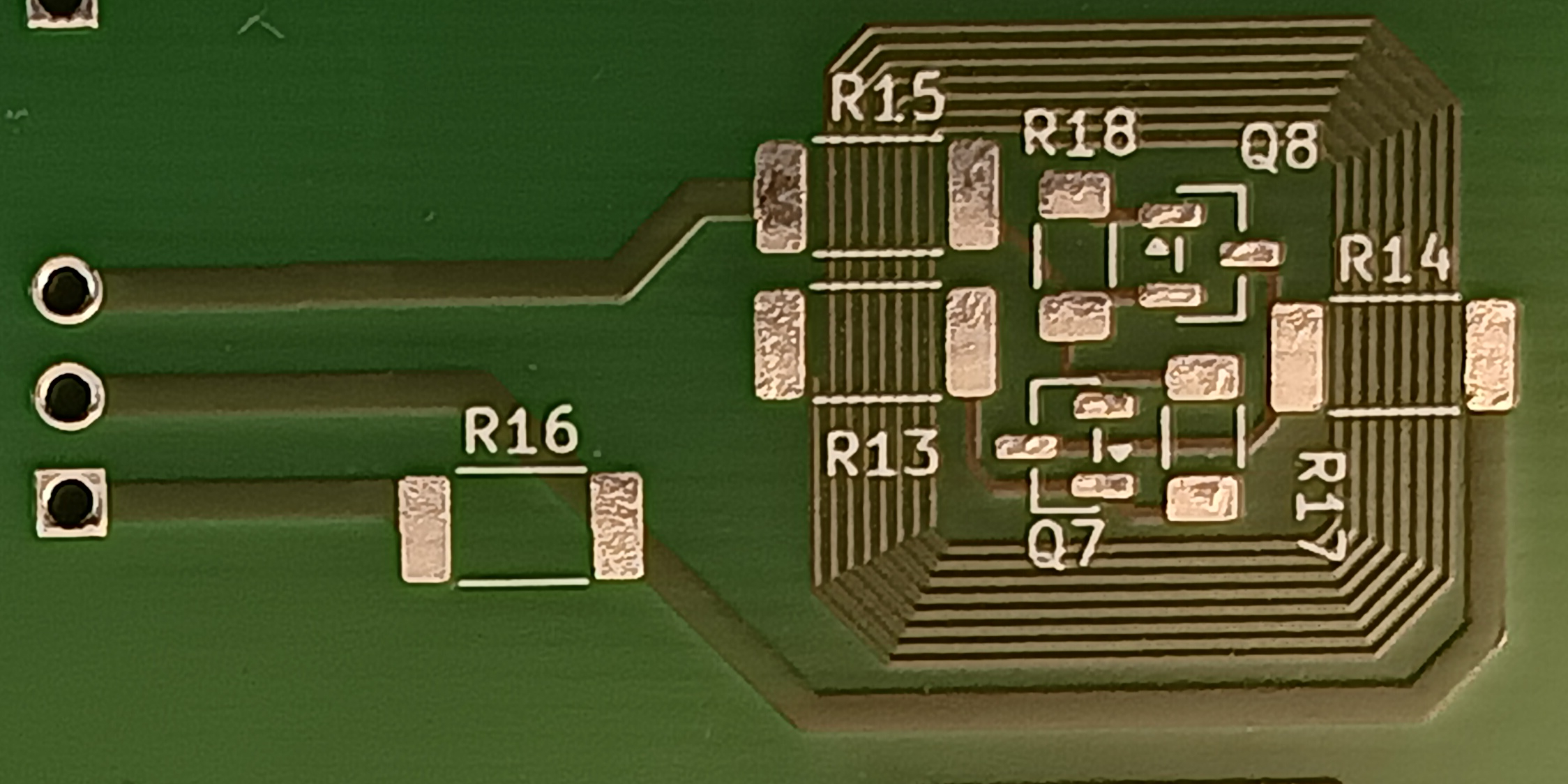}
\end  {center}
    \caption{Layout of transmitter (top side) and receiver (bottom side) for 2 layer PCB}
    \label{fig:basiclayout}
\end{figure}

  Using SOT23 devices, i.e. the 2N7002 and MMBT3904 for the transistors
  with 2010 sized resistors allows for the PCB layout shown in Fig.~\ref{fig:basiclayout}.
  An 8 pin DIP packaged IC is also in the picture as a reference scale.
  (NOTE: The 2N7000 equivalent SMD device 2N7002 ($<$ 200mA) should be
   used and not higher current, lower Ron variants as they will not operate 
   satisfactorily with the inductors shown here.)

  The coil is formed with 8 turns of an almost square coil of 200um wide track 
  and 200um track spacing. The length of the inner coil track is 9.7mm on 
  each side.  A 2 layer 1.6mm thick FR4 PCB with 35um copper thickness was
  used for the prototypes.  The overall coil area is about 16mm by 16mm.
  On the transmitter side, the two 2N7002 devices are placed inside the coil
  and 2010 SMD pads R9 and R12 are used to bridge across the coil to avoid the need 
  for vias (R12 is actually not needed as mentioned before). 
  R10 and R11 are short circuits (or 0R resistors) with R10 taking the 0V line into
  the circuit and R11 taking the outside of the coil back into the central circuit.

  The bottom side coil is identical to the top side coil and directly underneath it
  to maximize coil magnetic coupling.  The resistors R18 and R17 are 1206 size and
  together with SOT23 transistors Q8 and Q7, fit inside the coil interior.  Again
  bridging resistor pads of 2010 can be 0R resistors to connect across the coil without
  vias.  The pull-up resistor R16 is then outside the coil and can be part of the 
  circuit being driven in the application.

\section{Prototype Measured Performance}

\begin{figure*}[htb]
\begin{center}
  \includegraphics[width=0.75\textwidth]{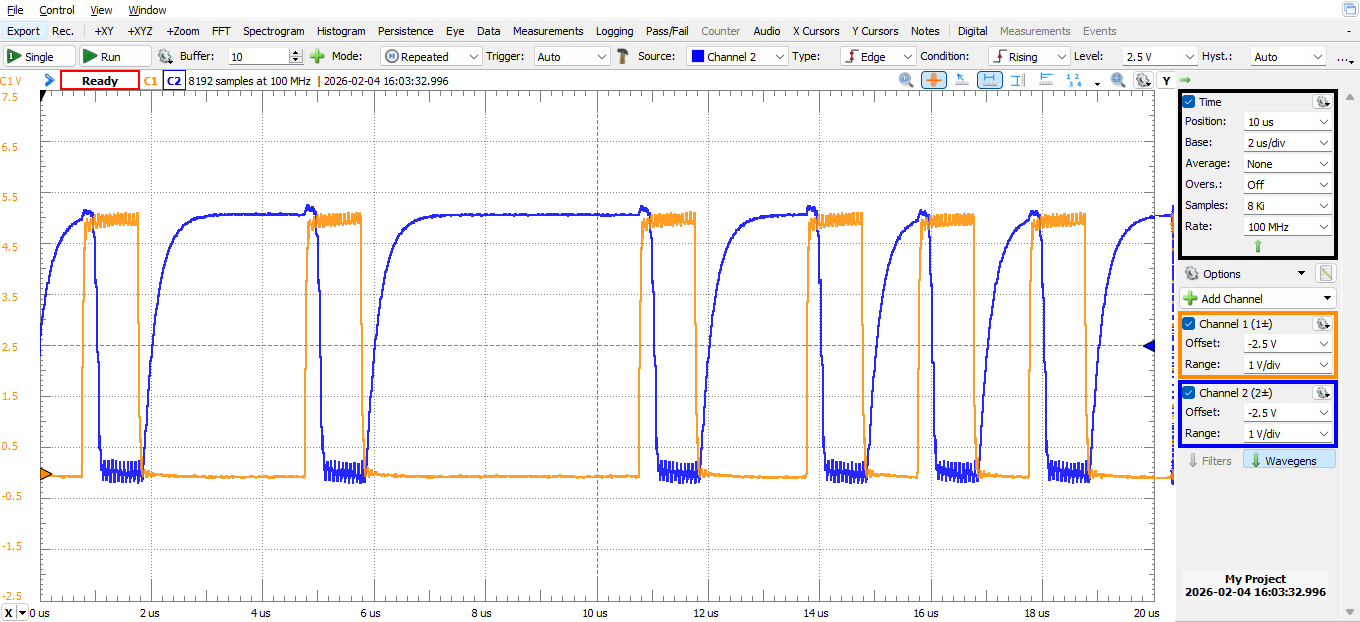}
\end  {center}
    \caption{Measured logic transfer for the prototype PCBs with a 1 Mbps NRZ random
    data signal}
    \label{fig:meas1operation}
\end{figure*}

  The measured performance of the circuit of Fig.~\ref{fig:basicschematic} implemented with the PCB layout of
  Fig.~\ref{fig:basiclayout} is shown in Fig.~\ref{fig:meas1operation}. 
    The orange trace (DX in Fig.~\ref{fig:basicschematic}) is the input 0/5V digital 
  signal from a 1 Mbps NRZ data source.   The blue trace shows the output (node DY)
  which is the secondary side data (inverted).  A logic propagation delay of around 200ns is 
  seen which should be satisfactory for a range of general purpose applications such as isolated UART, or SPI
  interfaces.

\begin{figure*}[htb]
\begin{center}
  \includegraphics[width=0.75\textwidth]{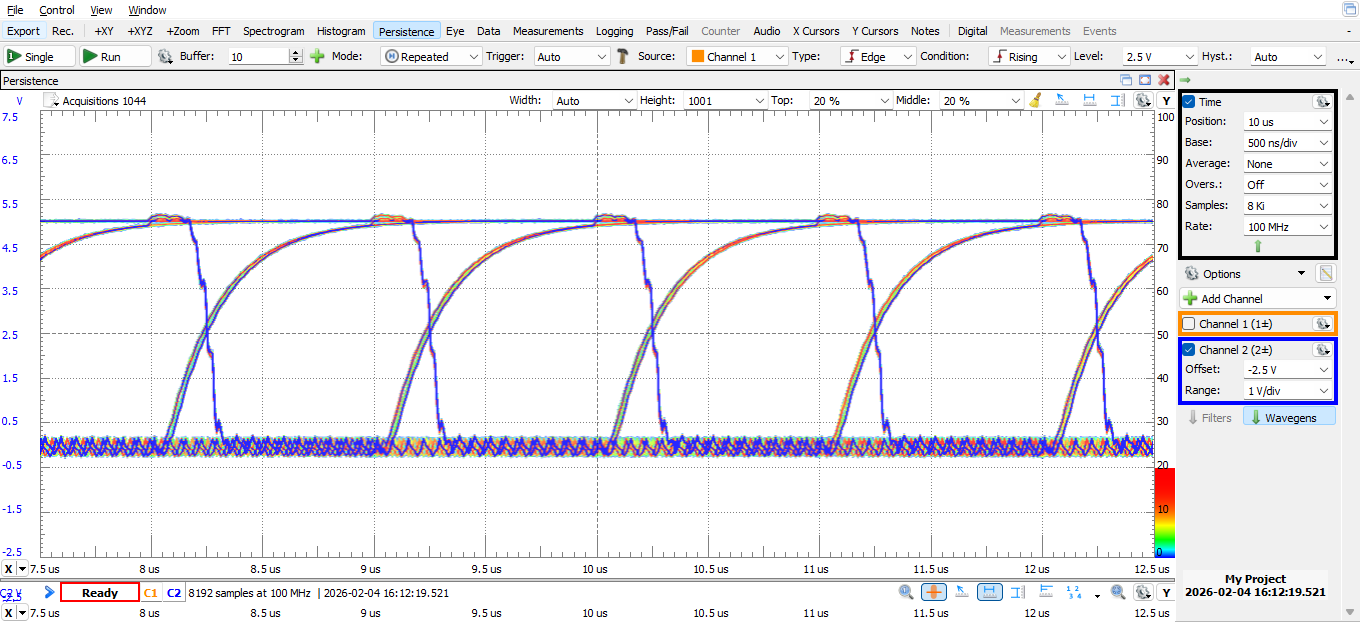}
\end  {center}
\caption{Corresponding Eye Diagram  for 1 Mbps NRZ random data}
\label{fig:NRZeye}
\end{figure*}

    Fig.~\ref{fig:NRZeye} shows the corresponding eye diagram which is a good visual representation of
the digital transmission ability of the circuit readily supporting a 1 Mbps data transfer rate.

The overall characteristics of the discrete digital ioslator circuit presented in this paper
are summarized in table~\ref{tab:isochars}.


   \begin{table*}[htb]
      \begin{center}
	 \caption{Overall Isolator Characteristics}
      \label{tab:isochars}
         \begin{tabular}{|c|c|c|c|c|c|c|}
            \hline
      Voltage &PCB Area   &    Power        &  Propagation  &  Data Rate  & Est Cost  \\
		 Isolation &&    Dissipation  &  Delay        &  Rate       & (incl PCB area)\\
            \hline
            \hline
		  &&   &       &                &     \\
		 $>$ 1kV  &3.4 cm $^2$    & TX 25mW (max)  & 200ns  &  $>$ 1 Mbps &   50 \textcent  \\
		  && RX  5mW (max)  &       &                &     \\
           \hline
         \end{tabular}
      \end{center}
   \end{table*}

\section{Low and High Side Signalling for Half Bridge Driver}

 One of the motivating applications for the circuit development is for signalling
 to the high and low side drivers for a half bridge power electronics circuit.
 When using opto-couplers for this application, it is common to connect the two LEDs
 in anti-parallel, so that applying a positive voltage turns one on with the other
 reverse biased, and a negative voltage turns on the other LED, but both LEDs cannot
 be simultaneously on, thus providing an inherent hardware lockout to prevent both 
 the high and low side drivers simultaneously conducting.

\begin{figure*}[htb]
\begin{center}
  \includegraphics[width=0.75\textwidth]{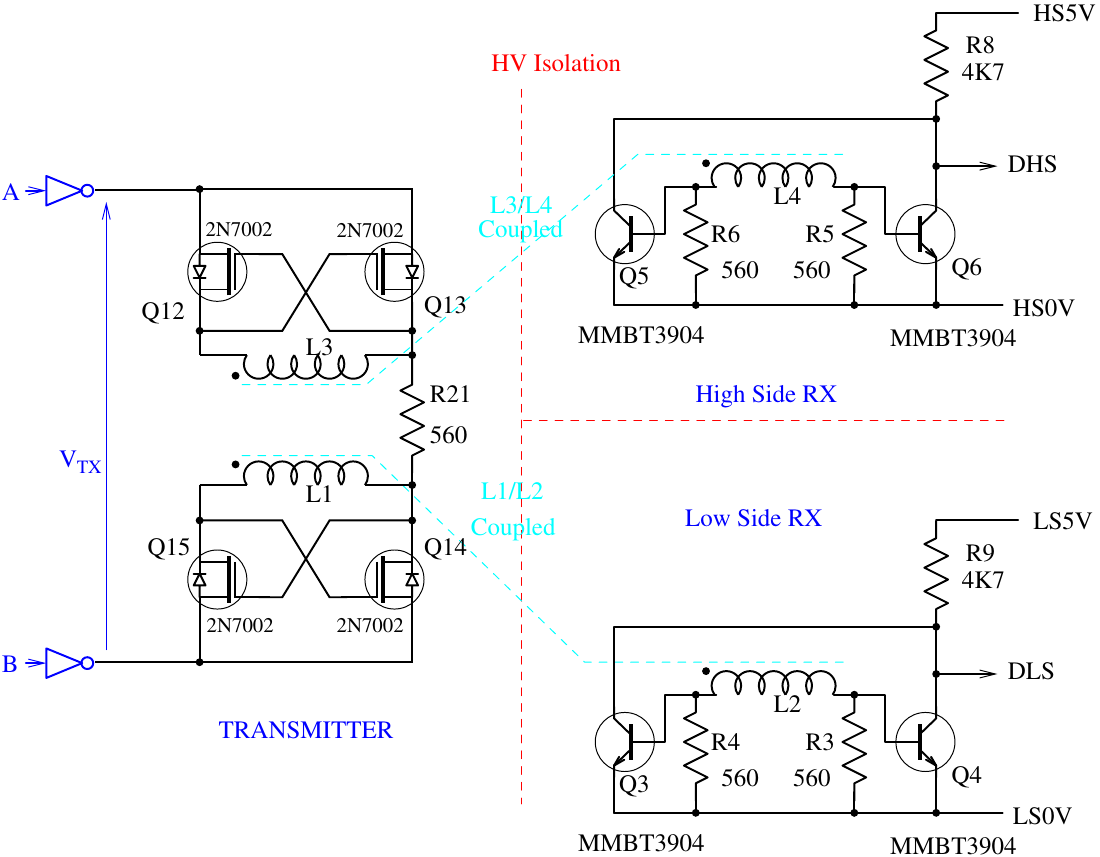}
\end  {center}
\caption{High/Low Side Isolation for Half Bridge Application}
\label{fig:hiloIsoDrv}
\end{figure*}

 Analogous to this, the circuit of Fig~\ref{fig:hiloIsoDrv} employs two digital isolators but
 combines the two transmitter oscillators into a single circuit, one 'on top' of the other.
 With the transistor polarities as shown and a 5V logic operation, consider A a logic 0
 and B a logic 1, then the voltage labelled $V_{TX}$ will be +5V and Q12 and Q13 will
 simply conduct through their internal source-drain diodes supplying power to the 
 oscillator consisting of Q14,Q15, L1 and R21 and coupling the oscillating current
 of L1 into the coupled coil L2 and pulling the low side driver signal {\tt DLS}
 low (to an active low power driver).  

 With A a logic 1 and B a logic 0, then $V_{TX}$ will be -5V and Q14 and Q15 will
 conduct through their internal source-drain diodes supplying power to the
 oscillator consisting of Q12,Q13, L3 and R21 and coupling these oscillating current
 of L3 into the coupled coil L4 and pulling the low side driver signal {\tt DHS}
 low (to an active low power driver). 

 With A and B both logic 1 or both logic 0, then $V_{TX}$ = 0 and the circuit is un-powered
 and neither driver activated.  Thus the circuit prevents both drivers being activated
 together and is suitable for the half bridge power electronics circuit configuration.

\begin{figure*}[htb]
\begin{center}
  \includegraphics[width=0.75\textwidth]{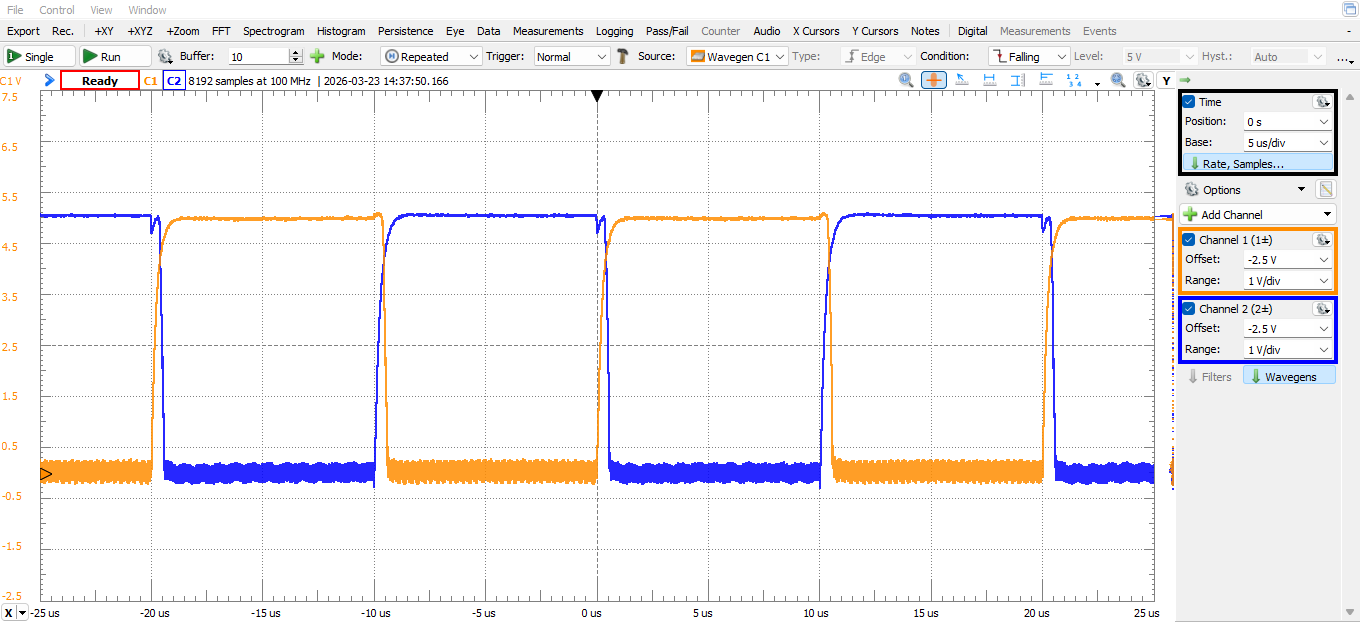}
\end  {center}
    \caption{High side, low side measured output with a 50 kHz $\pm$ 5V square wave input}
    \label{fig:HiLoIsoOp}
\end{figure*}

 Fig.~\ref{fig:HiLoIsoOp} shows the measured output for the circuit of Fig~\ref{fig:hiloIsoDrv}
 implemented on a PCB with similar layouts to those in Fig~\ref{fig:basiclayout}.
    The measurement shows clean logic waveforms with a natural dead time of about 300ns 
    (noting that the signals are active low).
 This level of performance is suitable for modest switching frequency power electronics 
 employing IGBT half bridge applications.

\section{Conclusions}

This article has described a simple discrete circuit digital isolator using commonly
available components that 
achieves isolation through the use of a PCB transformer with coils on opposite
sides of the board and relying on the FR4 board material for the voltage isolation.
Such a circuit can easily provide digital data rates of 1 Mbps or can be useful for
isolated signalling for power electronics applications such as the high side/low side
half bridge power circuit.

The key attribute of the design is the use of ubiquitous general purpose transistors
and no special purpose vendor specific parts, thus allowing for a design that supports
long life products in critcial applications where single supplier or vendor specific 
solutions and potential problems with obsolescence are of concern.

\begin{IEEEbiography}[{\includegraphics[width=1in,height=1.25in,clip,keepaspectratio]{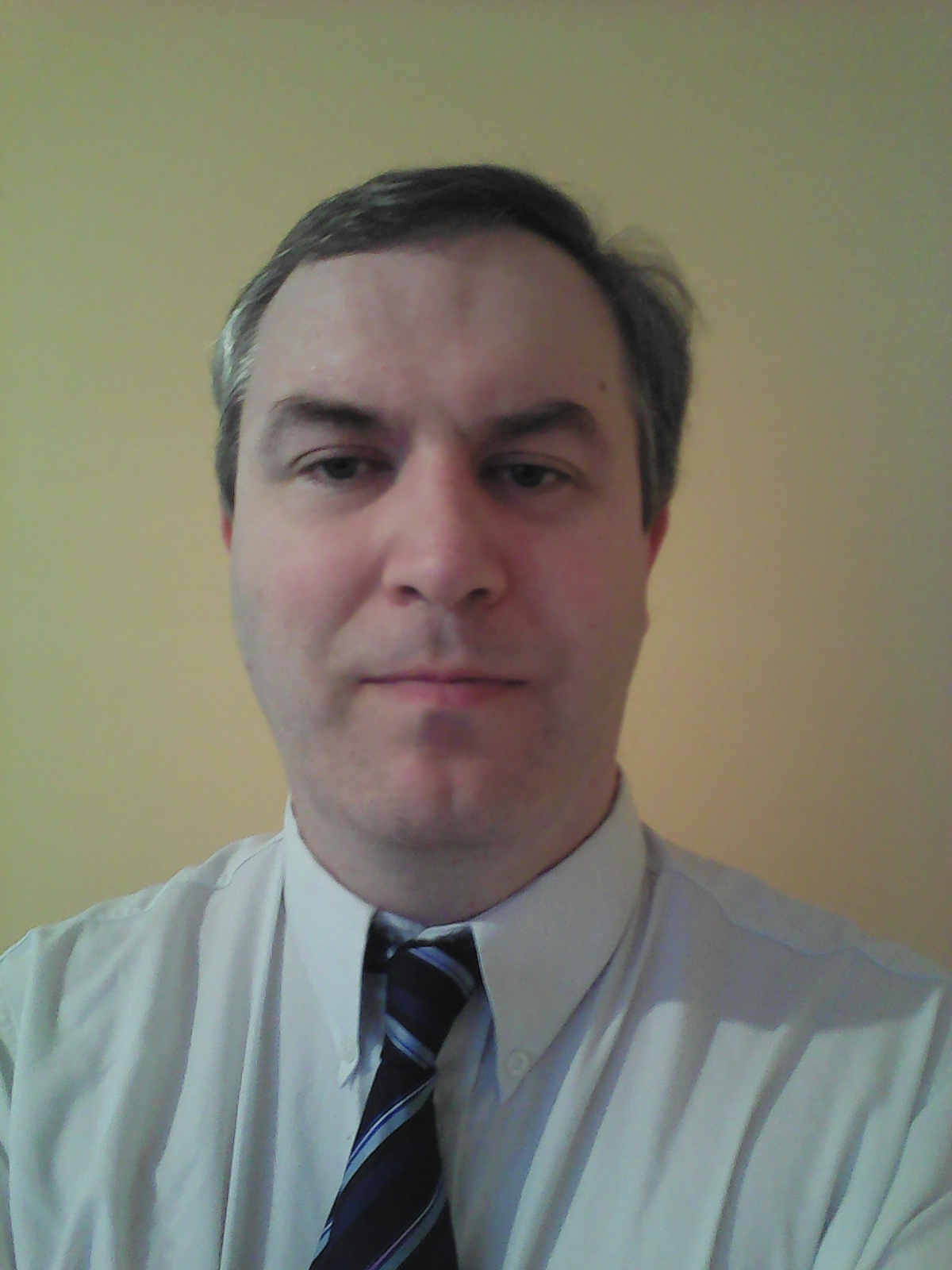}}]{Thomas Conway}
received the B.Eng. degree in electronic engineering and the Ph.D. degree in detection methods for magnetic recording channels from the University of Limerick, Ireland, in 1991 and 1996, respectively.

He joined Hewlett Packard CPB, Bristol, in 1991 and worked for two years on the design and development of tape drives for computer storage. From 1996 to 1998, he worked as a senior design engineer with Analog Devices B.V. Limerick, Ireland and Adaptec Inc., Longmont, CO, USA on disk drive read channels and controllers.

  He joined the ECE dept at the University of Limerick as a Lecturer in 1999 and lead the Limerick team in the European project on two dimension optical data storage, TWODOS, which was lead by Philips Electronics, NL with teams from TU Eindhoven, NL and Lancaster, UK.
   He was a Science Foundation Ireland funded investigator in the mid 2000s, successfully supervising 3 PhDs each going on to work with industry across Europe.

He also started and ran  a private consultancy business ALTCWY, providing consultancy to a range of companies both within Ireland and 
internationally.

In the 2010s, he turned his research to a life long interest in electric vehicles and sustainable energy. 

He has published extensively including 14 IEEE Transactions journals of which he is the primary author on 9, as well as being a named inventor on 9 patents.

\end{IEEEbiography}


\begin{thebibliography}{1}

\bibitem{ref:optoIso}
Broadcom. Optoisolation and 
Optical Sensor Products. Available at: https://docs.broadcom.com/doc/AV00-0254EN [[accessed 21 Mar 2026].

\bibitem{ref:AnalogDevicesADuM361N}
Analog Devices. ADuM361N: 3.0 kV rms 6-Channel Digital Isolator. Rev. D, accessed 21 Mar 2026. Available at: https://www.analog.com/en/products/adum361n.html.

\bibitem{ref:TIcapiso}
Texas Instruments. ISO73xx Series Capacitive Digital Isolators. TI Digital Isolator Design Guide (Application Note SLLA284). Available at: https://www.ti.com/lit/an/slla284g/slla284g.pdf [accessed 21 Mar 2026].

\bibitem{ref:Obsolescence00}
R. Solomon, P. A. Sandborn and M. G. Pecht, "Electronic part life cycle concepts and obsolescence forecasting," in IEEE Transactions on Components and Packaging Technologies, vol. 23, no. 4, pp. 707-717, Dec. 2000, doi: 10.1109/6144.888857.

\bibitem{ref:Obsolescence25}
S. Karaani, M. Zolghadri, M. Besbes, C. Baron, M. Barkallah and M. Haddar, "Systematic Analysis of the Links Between Obsolescence–Shortage and Reliability–Maintainability–Availability," in IEEE Access, vol. 13, pp. 88371-88389, 2025, doi: 10.1109/ACCESS.2025.3570107. 

\bibitem{ref:aircore1}
I. Boccato, R. La Rosa, I. Nikiforidis, P. D. Mitcheson, N. Aiello and C. Florian, "A Low-Profile GaN-Based Multi-MHz DC-DC Converter with an Air-Core PCB Transformer," 2025 Energy Conversion Congress \& Expo Europe (ECCE Europe), Birmingham, United Kingdom, 2025, pp. 1-6, doi: 10.1109/ECCE-Europe62795.2025.11238890.

\bibitem{ref:aircore2}
J. Sabate, E. Delgado and M. Harfman-Todorovic, "Gate Driver Power Supply for Medium Voltage SiC Mosfets with Air Core Transformer," 2022 IEEE Energy Conversion Congress and Exposition (ECCE), Detroit, MI, USA, 2022, pp. 1-6, doi: 10.1109/ECCE50734.2022.9948213.

\bibitem{ref:pcbspiralInd}
S. C. Tang, S. Y. Hui and H. S. . -H. Chung, "Characterization of coreless printed circuit board (PCB) transformers," in IEEE Transactions on Power Electronics, vol. 15, no. 6, pp. 1275-1282, Nov. 2000, doi: 10.1109/63.892842. 

\bibitem{ref:tiCreep}
SLUP421 – “Demystifying Clearance and Creepage Distance for High‑Voltage End Equipment”
Available at: https://www.ti.com/lit/pdf/SLUP421 [accessed 21 Mar 2026].

\end{thebibliography}
\end{document}